# Delayed Impact of Interdisciplinary Research

Yang Zhang[1], Yang Wang[1,a], Haifeng Du[1], and Shlomo Havlin[2,a]

[1] School of Public Policy and Administration, Xi'an Jiaotong University, Xi'an, 710049, China

[2] Department of Physics, Bar-Ilan University, Ramat-Gan 52900, Israel

[a] Correspondence and requests for materials should be addressed to Y.W. (email: yang.wang@xjtu.edu.cn) or to S.H. (email: havlin@ophir.ph.biu.ac.il)

## Abstract

Interdisciplinary research increasingly fuels innovation, and is considered to be a key to tomorrow's breakthrough. Yet little is known about whether interdisciplinary research manifests delayed impact. Here, we use the time to reach the citation peak to quantify the highest impact time and citation dynamics, and examine its relationship with interdisciplinarity. Using large scale publication datasets, our results suggest that interdisciplinary papers show significant delayed impact both microscopically per paper and macroscopically collectively, as it takes longer time for interdisciplinary papers to reach their citation peak. Furthermore, we study the underlying forces of such delayed impact, finding that the effect goes beyond the Matthew effect (i.e., the rich-get-richer effect). Finally, we find that team size and content conventionality only partly account for this effect. Overall, our results suggest that governments, research administrators, funding agencies should be aware of this general feature of interdisciplinary science, which may have broad policy implications.

Over the past centuries, scientific and technological advances have exploded to an unprecedented level, becoming a major fuel in today's economy [1-5]. The power of science has been confirmed by many examples from airplanes, vaccines, computers, and Internet, just name a few. As the Nobel Laurent Robert J. Shiller wrote, *"In the longer run and for wide-reaching issues, more creative solutions tend to come from imaginative interdisciplinary collaboration"*. Today, scientists increasingly work together on complex problems, ensuring interdisciplinary nature and higher impact outcomes and visibilities [6-10]. Also, interdisciplinary research is often considered as the key to solving societal problems [11]. The Human Genome Project, for example, has expanded into many disciplines and became a major interdisciplinary project that involves knowledge from various domains, providing new avenues for the advance of related disciplines, and increases the chance of curing common diseases such as cancer and diabetes. Understanding the impact of interdisciplinary research is crucial for managing a wide issues in science, from training junior scientists to supporting interdisciplinary research [12]. Despite potential societal impacts of interdisciplinary science, the question when does the actual major impact occur has rarely been discussed.

The availability of large scale datasets and advanced computational tools enable us to quantitatively understand this question [13-19]. On the one hand, the invention of the Science Citation Index, the development of the Internet search engines, and the extensive scientific collaborations enable scientific ideas to spread broader than ever, supporting the opinion that interdisciplinary research may have high and broad impacts [20, 21]. Recent studies confirm this hypothesis, suggesting that interdisciplinary research shows positive academic and technological impact [18, 22-27]. On the other hand, interdisciplinary research also faces numerous hurdles or paradoxes, including applying for funding [17, 28], establishing research centers [28], individual scientists conducting interdisciplinary research [29, 30], and obtaining prestige prizes [15-17, 28]. Such paradoxes indicate that it may take longer for interdisciplinary research to obtain its major impact because of the resistance of the classical communities [31].

Indeed, earlier research found negative correlations between paper interdisciplinarity and its short-term citations [18, 32]. Additionally, papers that were not recognized for a long time are more likely to be awakened by discoveries from other disciplines [33]. These views suggest plausible citation delay of interdisciplinary research. Yet at the same time, other observations indicate that there may not be such effects. For example, interdisciplinary research was shown to exhibit citation advantage regardless of time [22]. Also, disciplines of multidisciplinary nature tends to attract citations faster than traditional disciplines [34]. These divergent opinions suggest that the impact profile of interdisciplinary research remains unclear. Given the consequential nature of this question for funding agencies, individual researchers, scientific collaborations, as well as institutions or policies that support them, and building on recent substantial advances in understanding science [3, 4, 30, 35-42], we ask a fundamental yet unsolved question: does interdisciplinary research show delayed impact?

To offer quantitative answers to this question, we leverage the Microsoft Academic Graph (MAG) dataset [43], containing more than 200 million digital publications spanning more than 100 years. We extract publication date, scientific fields, journal, authors, author affiliations, citations and references for each paper. Overall, there are 37,808,446 journal articles till 2020. In this paper, we mainly focus on journal articles published before 2007 in order to study the long-term citation dynamics (see **Supplementary Material S1** and **Methods** for more sample details). We find that high interdisciplinary papers, on average, take longer time to reach their citation peak. Moreover, this effect is even larger for papers with extremely delayed citation peaks, and the citation peak of such papers is higher compared with low interdisciplinary papers. We also found that high interdisciplinary research published by prestige journals, prominent authors, and famous institutions also exhibits such delayed impact, suggesting that such effect is independent of reputations. Finally, our results suggest that organizational features and content conventionality are partly responsible for such delay.

## Results

A paper's reference list reflects its sources of input knowledge, and citation relationships between papers characterize knowledge diffusion [44]. This fact enables us to quantify paper interdisciplinarity using each paper's reference list. For individual papers, we thus quantify the extent to which each work integrates pervious wisdom, harnessing the *Rao-Stirling* diversity [45]. Specifically, the interdisciplinary indicator is defined as follows:

$$RS_d = \sum_{i \neq j} d_{ij} p_i p_j,$$

where $p_i$ and $p_j$ represent respectively the fraction of references in field $i$ and $j$, $d_{ij}$ represents the distance between field $i$ and field $j$ [23] (Detailed information is in **Methods**). This measure thus obtains a small value if a paper draws on knowledge from a set of similar disciplines or one single discipline (**Figure 1A, left**), and obtains a large value if a paper absorbs knowledge from various disciplines that are distant from each other (**Figure 1A, right**). Besides the *Rao-Stirling* diversity featured in the main text, we also utilize several other interdisciplinary measures (e.g., true diversity indicator [19, 46], and *DIV* indicator [19]) (see **Supplementary Material S2** and **Figure S2**).

First, we show an illustrating example of Price's seminal paper published in 1976 entitled *"A general theory of bibliometrics and other cumulative advantage process"* (which was among the first to quantify mechanisms that lead to fat-tail citation distributions) (see **Figure 1B** for its yearly citation dynamics) [47]. Despite of its substantial contributions, we observe that this paper gauged very little attention within 20 years after publication. After 2000, however, Price's discovery has suddenly attracted substantial citations, which probably due to the development of network science (**Figure S3**). Interestingly, we find that Price's paper is highly interdisciplinary (it lies in the top 1% of interdisciplinary level of all scientific articles published in 1976) (**Figure 1B, inset**). Indeed, this paper was inspired by a set of diverse scientific fields including statistics, data science, applied mathematics, library science, etc. The citation dynamics of this article vividly illustrates a common scenario in science, i.e., delayed recognition. As Garfield noted [48], *"Price suggests his paper is suffering delayed recognition because it involves a heavy mathematics discussion",* and this discussion raises our question: Do interdisciplinary papers universally show delayed impact?

To this end, we categorize all papers into three equal-sized subsets according to their interdisciplinary levels, i.e., high, medium, and low interdisciplinary research. After that, we calculate for each group the average yearly citations within 10 years after publication, finding that low interdisciplinary research systematically shows lower yearly citations compared to its counterparts, consistent with earlier studies [22]. Surprisingly, high interdisciplinary research tends to reach its citation peak substantially later than low interdisciplinary research, suggesting a delayed citation accumulation pattern (**Figure 1C**). The results are robust with respect to scientific fields and citation impact (**Figure S8**). We find that the time to reach the citation peak calculated from average yearly citation curves is, on average, 7 years for high interdisciplinary research compared to 5 years for medium interdisciplinary and 3 years for low interdisciplinary research (**Figure 1D**). To further support the significance of this difference, we employ a bootstrapping method, finding that the difference between the time to reach the citation peak is statistically significant (*p*-value < 0.001). These results suggest that macroscopically, i.e., collectively, high interdisciplinary research tends to follow substantial delayed citation accumulation patterns, which prompts us to ask a further question: Can we observe similar phenomena when considering microscopically (i.e., the citation peak for individual papers)?

To answer this question, we systematically investigate citation dynamics for individual papers. We define the peak time of a paper as the time to reach its citation peak after publishing, denoted by $T_m$. To ensure that the peak is sufficiently high compared to the background, the citations at the peak must exceed a threshold (we consider the threshold as the average citations plus two standard deviations for all yearly citations of the paper, shown by the horizontal dashed line in **Figure 2A**). Here, we ignore papers without citations or $T_m$, accounting for 30% of the sample. **Figure 2A** and **Figure 2B** show illustrations of papers. Specifically, **Figure 2A** shows a paper above its threshold of 11.33 citations at $T_m = 2$ years, whereas the paper in **Figure 2B** is above a threshold of 133.43 citations at $T_m = 45$ years. Note that the one with larger $T_m$ shows substantially high interdisciplinarity, i.e., the *Rao-Stirling* diversity is 126.9% higher than the paper with the smaller $T_m$.

We find that high interdisciplinary papers show substantially higher average $T_m$ than low interdisciplinary papers, as $T_m$ increases monotonically with paper interdisciplinarity (**Figure 2C**). Specifically, as papers interdisciplinary level increases from bottom 5% to top 5%, their average $T_m$ increases from 7.1 years to 8.6 years, by a factor of 20% (*t*-test *p*-value < 0.001). We further group the papers by different scientific fields, and find similar results (**Figure 2D**). Specifically, high interdisciplinary research in medicine, chemistry, biology and physics (these four fields account for over 70% of all papers) waits for 2.7%, 7.2%, 29.2%, 21.6% more time to reach the citation peak compared to low interdisciplinary research, respectively (*p*-values for all cases are smaller than 0.001). These results support and quantify the hypothesis that high interdisciplinary research, on average, shows more delayed impact, whereas low interdisciplinary research tends to attract attention faster.

This pattern is amplified when we focus on papers with extremely delayed citation peaks (top 5% $T_m$ of papers published in the same year). Papers in the top 5% interdisciplinary level is almost 44% more likely to exhibit extremely high $T_m$. By contrast, low interdisciplinary research is less representative among papers with extremely high $T_m$ (**Figure 2E**). Comparing high and low interdisciplinary research, we find that high interdisciplinary papers is 62% ((1.44 − 0.88)/0.88) more likely to show extremely delayed impact compared with low interdisciplinary research (**Figure 2E**). Interdisciplinary research also shows higher maximal impact (denoted by $C_m$). We plot the average $C_m$ as a function of paper interdisciplinarity for papers with top 5% $T_m$, finding significant positive association (**Figure 2F**). Specifically, among all papers with top 5% $T_m$, top 5% interdisciplinary research, on average, attracts 54.1% more citations at year $T_m$, compared to bottom 5% interdisciplinary papers (*t*-test *p*-value < 0.001). These results indicate that interdisciplinary research not only shows delayed impact, but also has significantly higher maximal impact.

We repeat our analysis in several directions to further support our observations. We define $T_m$ using a 3-year-moving average of individual paper yearly citations in order to eliminate possible noise, and without any threshold, finding similar results (**Supplementary Material S4.3 and Figure S9**). We repeat all analysis using only 10 years or 20 years citation windows, and find that window size does not affect our results (**Supplementary Material S4.3 and Figure S9**). We also directly compare the distribution of $T_m$ for low, medium, and high interdisciplinary research, finding that these distributions show significant differences (**Supplementary Material S4.4 and Figure S10**). We repeat the analysis for papers with similar number of citations received within 10 years after publication (i.e., $C_{10}$) across different time periods, finding that the average $T_m$ of high interdisciplinary research is consistently higher using different samples (**Supplementary Material S4.4 and Figure S11**). Finally, we directly use the beauty index [33] and the impact time [35] to quantify the citation dynamics, obtaining similar results (**Figure S12, Figure S13**). Together, we used large-scale scholarly datasets to demonstrate the consistent positive relationship between paper interdisciplinarity and the time to reach its citation peak, both macroscopically and microscopically.

Next we ask, can the above positive association be attributed to other factors? To this end, we leverage the fixed effect regression with the predicted variable $T_m$, while controlling for cofounding factors. **Figure 3A** shows the predicted margins of $T_m$ conditional on both, paper interdisciplinarity and the number of references, while controlling for team size, author academic impact, author academic age, institution prestige rank, scientific field, publication time, publication venue (detailed information regarding the regression settings can be found in the **Methods**). We find an intriguing pattern that papers of high interdisciplinary (top 30% interdisciplinary papers) are characterized by larger $T_m$ compared to others regardless of their number of references. Specifically, high interdisciplinary papers take, on average, 8.6% longer time to reach their citation peak while adjusting possible influencing factors (see **Table S1** for the full regression table). High interdisciplinary research takes longer time to reach the citation peak regardless of its ultimate academic impact (**Figure 3B**). We find that high impact work with high interdisciplinary nature takes longer to reach its citation peak compared to its counterparts with low interdisciplinary level. Specifically, high impact papers in the top 5% interdisciplinary nature need to wait 9.8% more time compared to those in the bottom 5% interdisciplinarity with similar citations, whereas for low impact works the slope flattens (**Figure 3B**). We further evaluate the predicted margins of $T_m$ separately for medicine, chemistry, biology and physics, finding that these patterns are stable across these four major scientific fields (**Figure 3C**). Among these four fields, we find that the delayed impact of interdisciplinary research is more prominent in biology and physics.

We further support our findings by considering sleeping beauties in science, which corresponds to delayed recognition. Specifically, we compute directly the beauty index (i.e., $B$ index) for papers with different interdisciplinary levels (see details in the **Methods** and **Supplementary Material S3.2**) [33]. Large beauty index corresponds to longer sleeping duration and higher awakening intensity. **Figure 3D** shows the predicted margins of a paper's $B$ index as a function of different interdisciplinary levels, while we adjust for other factors using regression analysis (**Table S1**). We find again that high interdisciplinary papers tend to have higher probability to become sleeping beauties across various time periods. Specifically, in 1970s, top 10% interdisciplinary papers are 25% more likely to become sleeping beauties than the bottom 10% interdisciplinary counterparts. The slope flattens in recent years as $B$ index tends to be small gradually with respect to short citation windows. This observation is in line with the fact that papers with high $B$ index are more likely to attract citations from foreign disciplines [33].

We perform additional analyses to support our findings. First, we investigate the association between paper interdisciplinarity and $T_m$ for each year by conducting separate regressions (**Figure S15**), and use alternative interdisciplinary indicators (**Figure S14A**), finding similar results. To further study the dynamical patterns underlying individual paper citation dynamics, we fit each paper's citation dynamics to a generative model [35] (detailed model descriptions are shown in **Supplementary Material S3.1**). We then replace $T_m$ with relevant parameters in the model, and conduct the regression analysis, finding consistent results (**Figure S14B**). We also directly calculate individual paper's impact time measured as the time to arrive a paper's half citations (**Supplementary Material S5.1** and **TableS1**). To rule out the possibility that unobservable features for individual scientists affect our results, we add individual author fixed effect in the regression analyses, and we again find robust results (**Table S2**). Finally, Poisson regressions yield similar results (**Table S3**).

To understand the potential forces that yield our findings, we theorized such delay in relevant fields. Inspired by the literature on the sociology of science, we first investigate whether delayed impact pattern of interdisciplinary research would be affected by the Matthew effect, which suggests the rich-get-richer effect in science [49]. Several caveats we observed are intriguing. First, we find that $T_m$ decays dramatically from 9 years to less than 7 years as a function of the journal impact factor, suggesting that papers published in high impact venues attract academic attention significantly faster than their counterparts in modest impact venues (**Figure 4A**). Also, **Figure 4B** shows that research written by prominent scientists attracts approximately 4.8% faster than research written by ordinary researchers, consistent with the fact that early citation premium probably rooted in the reputation or status of authors [50]. We find that the positive gap of $T_m$ between high and low interdisciplinary research goes beyond the Matthew effect, as the gap remains significant regardless of venue and author impacts (**Figure 4A**, and **Figure 4B**). More precisely, high interdisciplinary papers published by prestige journals and prominent scientists, on average, wait for roughly 3% longer time to arrive to the citation peak, compared to low interdisciplinary papers with similar characteristics. We also use institutional rank as another proxy for reputation. Specifically, we match the institutions with the U.S. News Rank and use the lowest rank (low rank corresponds to high status) of a paper's institutions to quantify its reputation. We find that the difference of $T_m$ between high interdisciplinary research and its counterparts is stable, i.e., high interdisciplinary research by prestige institutions takes 5.5% longer time to reach the citation peak compared to low interdisciplinary research by institutions with similar reputation (**Figure 4C**). Together, these results are line with Price's example, which implies that high interdisciplinary research from high status journals/researchers/institutions also exhibits delayed impact compared to low interdisciplinary research of similar features.

Motivated by the literature on the team science, we here study the organizational features [51]. To this end, we look into whether the $T_m$ gap between high and low interdisciplinary research varies across different team sizes. For papers of small teams (less than 7 members), high interdisciplinary research still shows delayed impact compared to low interdisciplinary research, by 2.9% (**Figure 4D**). Interestingly, we do find indications that our finding of difference of $T_m$ between high and low interdisciplinary research shrinks for large teams, which collides with conventionality as teams often publish articles through incorporating novel ideas while not abandoning conventionality [36]. In light of this finding, we therefore estimate the content conventionality of each paper using journal pairs within the reference list [36], finding that combining high conventional wisdom shrinks the gap of $T_m$ between high and low interdisciplinary research (see **Supplementary Material S6.3** for more details). We further test whether conventionality acts as a possible mechanism why interdisciplinary science exhibits delayed impact. We find that interdisciplinary science tends to be unconventional, and the partial correlation between interdisciplinarity and $T_m$ shrinks when we control for paper conventionality (**Supplementary Material S6.3**). Together, these findings are consistent with what Cole has previously written [52], *"… the delayed recognition was primarily the result of content rather than the author's prestige"*.

## Conclusion

Despite intensive efforts in understanding interdisciplinary science and scientific impact, little study has focused on citation accumulation patterns for interdisciplinary research. Here, instead of looking into the static citations for various citation windows [18, 32], we focus on, both microscopically and macroscopically, the time to reach the citation peak. We find that interdisciplinary research exhibits delayed impact, as it takes longer time to reach the citation peak. Furthermore, the difference of the time to reach citation peak between high and low interdisciplinary research in prestige venues, or by prominent authors and famous institutions remains stable, suggesting that our findings are independent of the Matthew effect. Additionally, we find that the difference becomes much smaller for papers written by large teams or through combining high conventional wisdom, which collides with the claim about innovative strategies of large teams without giving up conventionality [36]. Our paper contributes to current discussions on complex relationships between interdisciplinarity and citation dynamics, while considering the different underlying forces.

The uncovered delayed impact of interdisciplinary research may have multiple reasons. We find indications that combining unconventional wisdom partly explains the effect. That's to say, interdisciplinary research that absorbs wisdom spanning diverse disciplines may be far away from the classical community, leading to delayed impact. Empirical analysis also finds that the gap of $T_m$ between high and low interdisciplinary science only occurs for unconventional papers (**Figure S16B**). Additionally, as interdisciplinary research often has broad impact [23] and such impact in foreign fields often takes longer time, the knowledge diffusion may also account for such delay. Together, we hypothesize that these two forces both play a role in this complex relationship between interdisciplinarity and time to reach the citation peak. Crucially, we find that our results hold true regardless of the underlying processes. Note that novel research also exhibits delayed recognition [53], and the overlapping between interdisciplinarity and novelty raises noticeable discussion that is beyond the scope of our paper [54].

Taken together, given the fact that interdisciplinary research is often considered as the space for innovation research that is the key to future economic growth, policy makers who push interdisciplinary research should notice such delayed impact. Specifically, interdisciplinary research may show substantial disadvantages using bibliometric indicators such as the journal impact factor, or essential science indicators that explicitly use very short citation windows. Funding agencies that extensively use bibliometric indicators need to develop or refine current evaluating system to foster interdisciplinary research. Moreover, for institutions that hire or nurture junior scientists engaging in interdisciplinary research, our results also provide empirical insights. Namely, too much focus on short-term citations may disincentive interdisciplinary research. Simultaneously, interdisciplinary researchers attain better funding performance in the long run [55], which depicts long-term advantage propensity of conducting interdisciplinary research. We advocate the awareness of such bias, especially for research administrators [53]. Finally, our analysis indicates that such delayed impact is an inherent nature of knowledge production process such as scientific paradigms [31] or knowledge diffusion across scientific domains [56], which may have long lasting effects on individual scientists, funding allocations [17], and prize systems [16].

## Methods

**The Microsoft Academic Graph.** The primary data source of this paper is the Microsoft Academic Graph (MAG) database, which covers more than 200 million digital publications. To study the association between interdisciplinarity and time to reach the citation peak, we extract, from the MAG data, each paper's publication time, scientific fields, references, forward citations, publication venues, authors, and institutions. We mainly focus on journal articles published before 2007 to ensure sufficient citation window length. In the sample, 94.5% papers cite at least two references with scientific field information. Finally, we end up with 16,548,023 papers, ranging from the year 1970 to 2007. Scientific field information in MAG dataset is grouped into a five-level hierarchical tree and each field represents a specific scientific discipline or topic. Here, in order to define a paper's interdisciplinarity, we consider 295 level-1 fields to which each paper belongs, and these 295 level-1 fields belong to 19 level-0 scientific disciplines. For instance, data science, artificial intelligence, machine learning (level-1 fields) all belong to "computer science" (level-0 field).

**The Global Universities Ranking Dataset.** To quantify institution prestige, we collect data from the U.S. News best global universities rankings. We accessed the data on August, 1st, 2021. This ranking evaluates academic performance and reputation of more than 1,500 universities around the globe, providing credible information for young students who look for education abroad. Specifically, U.S. News uses 13 weighted indicators to calculate university rankings, including reputations (25%), bibliometric indicators (65%), scientific excellence (10%), etc. For instance, bibliometric indicators include publications (10%), percentage of total publications that are among the top 10% of the most cited papers (10%), etc. There are 1,748 schools, which are ranked within top 1,500 universities. We manually match the U.S. News data with MAG institution information, to obtain institution rank for our sample. Moreover, if an institution is not in the list, we label it "no rank".

**Quantifying paper's interdisciplinarity.** Here, we mainly employ the commonly used Rao-Stirling diversity in order to quantify paper interdisciplinarity [18, 23, 24, 45]. In line with former research [23], the procedure of calculating interdisciplinarity is as follows. For each paper, we denote $\alpha$ to a vector $d_\alpha = (d_1, d_2, ..., d_k)$ based on its scientific fields, where $d_k$ is a field and $k$ is the number of fields in the MAG data (i.e., 295 level-1 fields). Thus the $i_{th}$ element of paper vector $\vec{a}$ is given by

$$a_i = \sum_{\alpha=1}^{n} \sum_{\beta=1}^{k} \frac{\delta(d_\beta, i)}{k},$$

where $\delta$ is the delta function with 1 when $d_\beta$ is the same as the $i_{th}$ scientific field, $n$ is the total number of references. Moreover, we normalize each paper vector in order to calculate the *Rao-Stirling* indicator [45]. Next, we define the field vector by aggregating vectors $\vec{a}$ over articles within the discipline, i.e., $v_i = \sum_{a \in V} a$. As we discuss next, it provides insight on the observed degree of disciplinary closeness. Generally speaking, as **Figure S2A** shows, the distance between fields within the same discipline is smaller, suggesting that knowledge flow in the same discipline is more frequent than knowledge flow across distinct disciplines. For example, the knowledge combined between data science and statistics physics is more related to each other than the knowledge captured between nuclear physics and biochemistry. The distance between field $i$ and field $j$ is defined using the cosine similarity, as described above:

$$d_{ij} = 1 - \frac{v_i \cdot v_j}{|v_i| \cdot |v_j|}.$$

Finally, for each paper $d$, we define its interdisciplinarity using the *Rao-Stirling* diversity:

$$RS_d = \sum_{i \neq j} d_{ij} p_i p_j,$$

where $p_i$ and $p_j$ represent the fraction of references in field $i$ and $j$, respectively. The calculation of the *Rao-Stirling* diversity is illustrated in **Figure 1A**. And the alternative interdisciplinary indicators are described in **Supplementary Material S2**.

**Quantifying citation dynamics.** Here, besides our definition of $T_m$ as illustrated in **Figure 2**, we also harness several methods to quantify citation dynamics of individual papers. One is a mechanic model, i.e., the WSB model (see details in **Supplementary Material S3**). Beyond the WSB model, it is well believed that a handful of articles attract substantial attentions many years after publication, and these kinds of articles cannot be captured by the WSB model as most of such articles do not follow rise-and-fall citation dynamical pattern. To this end, we harness the "sleeping beauty index ($B$ index)" to quantify citation dynamics of such awaken articles [33]. The $B$ index is determined by the duration of sleeping and the awaken intensity. More specifically, awaken intensity represents the strength of suddenly receiving substantial attention from the situation where few citations occur during initial decades for one article. The $B$ index of paper $d$ with at least one citation is defined as follows,

$$B_d = \sum_0^{t_m} \frac{\frac{C_m - C_{t_0}}{t_m} \cdot t + C_{t_0} - C_t}{\max\{1, C_t\}},$$

where $C_t$ is the number of citations at year $t$, $C_m$ and $t_m$ are the maximum number of citations and its corresponding year, respectively. **Figure S5** shows the graphical representation of calculating $B$ index till time $t_m$.

**Regression setting.** To further eliminate the effect of the confounding factors, we apply ordinary least squares regressions to study the association between paper interdisciplinarity and citation dynamics. The regression model is shown as follows,

$$y_d = \beta_i Inter_d + \beta_T \ln(T_d) + \beta_R \ln(R_d) + \beta_{af} A_{fd} + \beta_{al} A_{ld} + \beta_{aa} A_{ad} + \beta_{hf} H_{fd} + \beta_{hl} H_{ld} + \beta_{ha} H_{ad} + \sum_r \beta_r R_{rd} + \sum_j \beta_j J_d + \sum_y \beta_y Y_d + \sum_f \beta_f F_d + \epsilon_i.$$

The dependent variable $y_d$ represents the time to reach the citation peak for individual papers. Specifically, we use $T_m$ in the main text, and $u_d$ from the WSB model or the $B$ index. To check the robustness of our regression results, we also use the impact time, approximated as the characteristic year when an individual paper reaches its half total citations after publication till 2018. The predictor of interest is interdisciplinarity, i.e., the Rao-Stirling diversity in the main text. To consolidate the robustness of the results, we also use alternative interdisciplinary indicators as described in **Supplementary Material S2**. To ensure that our results are not affected by other factors, we explicitly control the following variables: 1) $T_d$ is team size. It measures the number of authors of a paper, and we transform the variable to its logarithm because of its fat-tail nature. 2) $R_d$ is the number of references. We also convert this variable into its logarithmic form. It is assumed that the more literatures a paper cites, the more easily it gets attention. 3) $A_{fd}$ is academic age of the first author. It measures the number of years from the first publication time till paper $d$'s publication time. Academic age partly indicates an author's academic experience. 4) $A_{ld}$ is academic age of the last author. It measures the number of years from the first publication time till paper $d$'s publication time. 5) $A_{ad}$ is average academic age of a team. It measures average number of years since the first publication till paper $d$'s publication time over all team members. 6) $H_{fd}$ is h-index of the first author. It measures the impact of the first author till paper $d$'s publication time. 7) $H_{ld}$ is h-index of the last author. It measures the impact of the last author till paper

$d$'s publication time. 8) $H_{ad}$ is average $h$-index of a team. It measures average impact of a team using the similar setting as $H_{fd}$. 9) $R_{rd}$ is fixed effect of the highest institution prestige as indicated by the U.S. News Ranking. This group includes 9 dummy variables for institution rank: [1, 10], [11, 20], [21, 40], [41, 80], [81, 160], [161, 320], [321, 640], [641, 1499] and "no rank". "No rank" contains institutions which could not be recognized by the U.S. News Ranking. Higher rank corresponds to lower prestige. 10) $J_d$ is fixed effect of publication venue. The publication venue of a paper is extracted from journal id in MAG dataset. 11) $Y_d$ is fixed effect of publication year. 12) $F_d$ is fixed effect of academic fields. The academic fields are indicated in level-1 fields from MAG dataset, arriving at 295 fields. For all regression analysis in this paper, the standard errors are clustered at individual journal level.

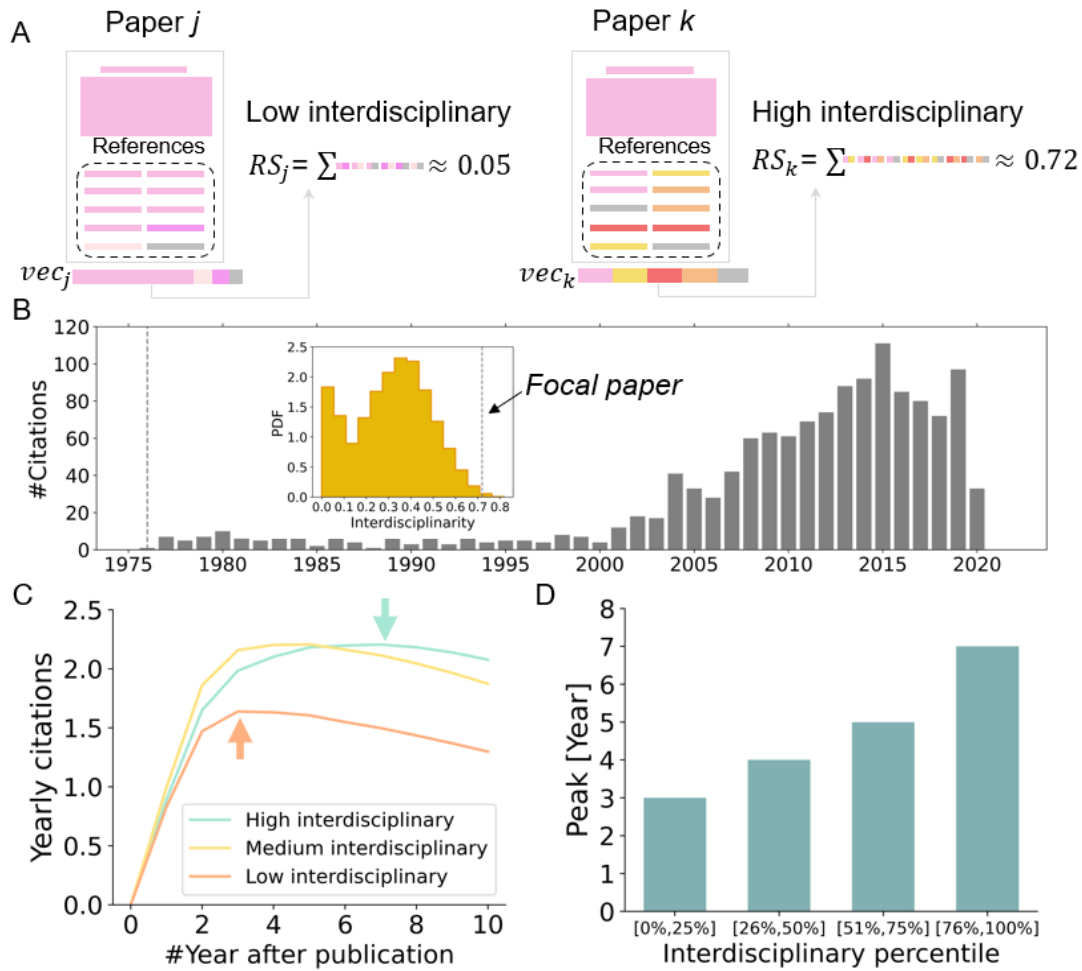

**Figure 1. Quantifying interdisciplinary nature and its relationship with citation accumulation patterns.** (A) An illustration of the interdisciplinary measure. The interdisciplinary measure quantifies how a paper is broadly inspired based on its reference list. Two papers could belong to the same scientific field, but absorb knowledge from several scientific fields with different distributions. An increasing value indicates broader inspiration integration, charactering with higher diversity of scientific fields in the reference list. (B) Citation history of the paper entitled "*A general theory of bibliometrics and other cumulative advantage process*" by Price; Inset figure shows that it lies in the top 1% interdisciplinary measure of papers in the same year. (C) Yearly average citations of all papers, i.e., macroscopic view, within 10 years after publication for high, medium, and low interdisciplinary research. It reveals a systematic delayed citation accumulation pattern, particularly the year of the peak value, for high interdisciplinary research compared to medium or low interdisciplinary research. (D) The peak time as a function of interdisciplinary percentile. We calculate peak time through the curves of yearly average citations of all papers within 10 years after publication for research belonging to different interdisciplinary percentiles. The difference is statistically significant (*p*-value < 0.001). The error bars are not shown since they are too small.

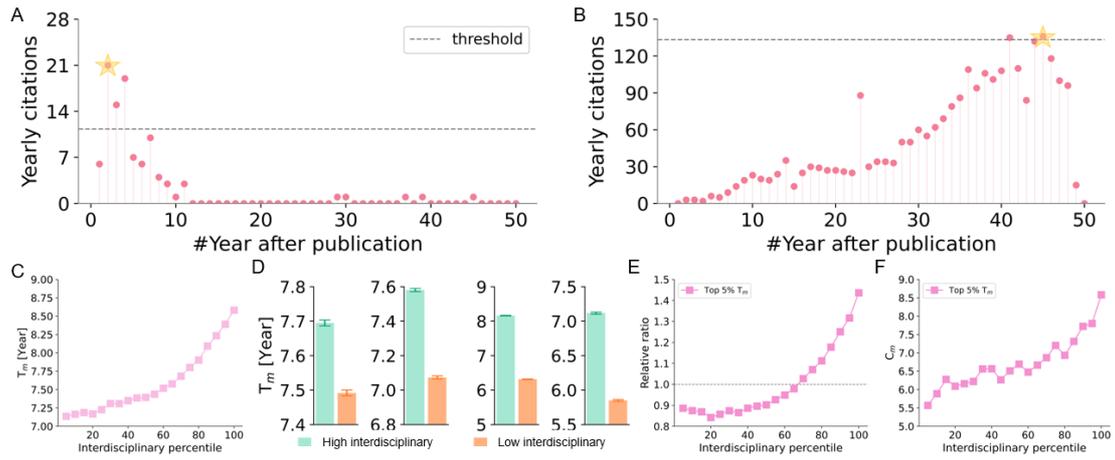

**Figure 2. Delayed impact of interdisciplinary research based on individual papers microscopically view.** (A-B) Illustration of $T_m$ for individual papers. The yellow star represents the time to reach the citation peak of the focal paper, where yearly citations need to exceed the threshold (average citations plus two standard deviations for all yearly citations of the focal paper, shown by the horizontal dashed line). We demonstrate (A) a paper with low interdisciplinary nature (i.e., Rao-Stirling diversity is 0.18) has $T_m$ = 2 years, whereas the other (B) with high interdisciplinary nature (i.e., Rao-Stirling diversity is 0.59) has $T_m$ = 45 years. (C) For individual papers, the average $T_m$ as a function of paper interdisciplinarity. Average $T_m$ increases by 20% from 7.1 years to 8.6 years as paper interdisciplinary level increases from the bottom 5% to the top 5%. (D) Comparing average $T_m$ for high and low interdisciplinary research across medicine, chemistry, biology, physics with error bars representing standard errors. (E) The same as in (A) but for extremely high $T_m$ papers. Relative ratios compare the observed fraction of papers of certain interdisciplinary levels with extremely high $T_m$ against the constant baseline of 5%. We find only 4.4% papers with extremely high $T_m$ are contributed by bottom 5% interdisciplinary papers and 7.2% contributed by top 5% interdisciplinary papers. (F) For extremely high $T_m$ papers, average $C_m$ as a function of interdisciplinary percentile.

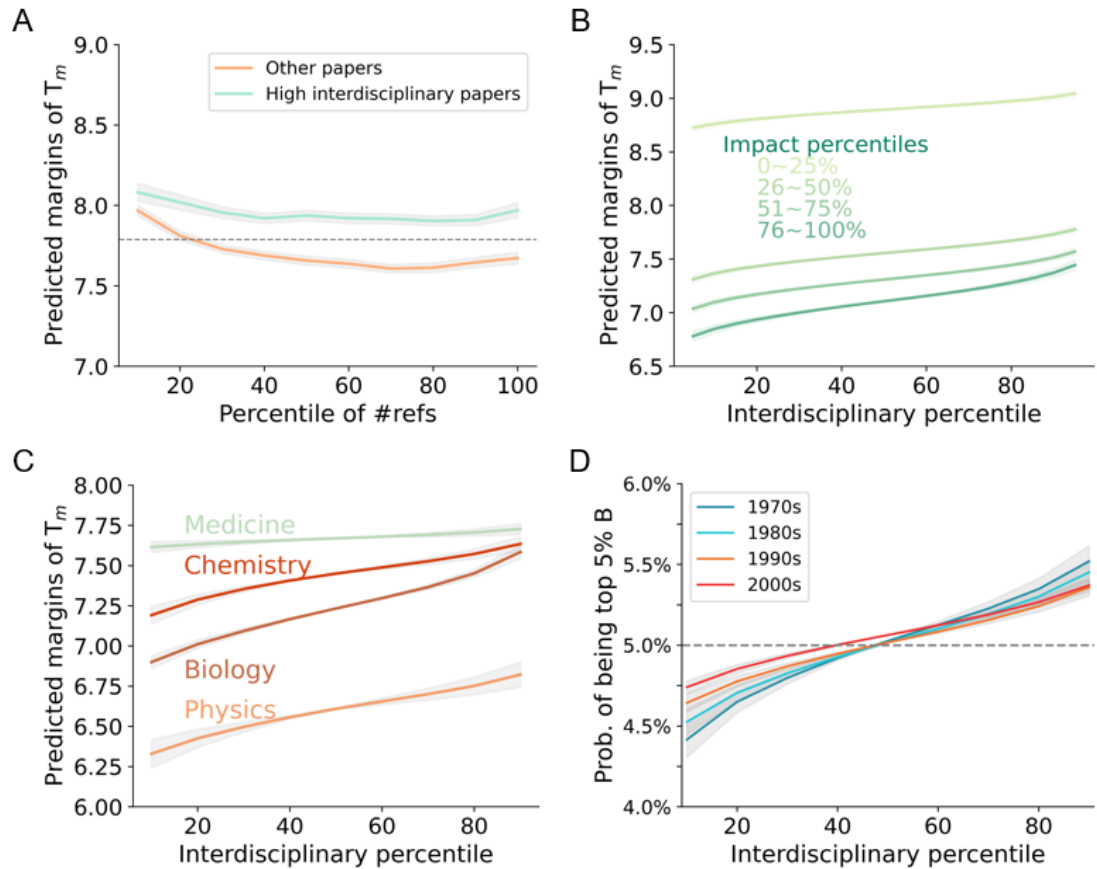

**Figure 3. Delayed impact of interdisciplinary research across impact levels, scientific fields and time periods.** (A) The predicted margins of $T_m$ conditional on the number of references and paper interdisciplinarity, while we adjust for possible confounding factors in the regression analysis. We group papers with the top 30% interdisciplinary nature as high interdisciplinary research, whereas the remaining papers are categorized regular research. The graph shows that high interdisciplinary research has larger and stable $T_m$ above the base line (dashed line) across different number of references, and likewise, compared to regular research. (B) The predicted margins of $T_m$ as a function of interdisciplinary percentile across various impact levels. Curves are colored by different impact percentiles (i.e., the number of citations captured within 10 years after publication). The predicted margins of $T_m$ increases faster for papers of higher impact percentiles. (C) The predicted margins of $T_m$ as a function of interdisciplinary percentile across medicine, chemistry, biology and physics. (D) The predicted margins of the probability of being sleeping beauties (i.e., the probability of being the top 5% $B$ index in the same year) as a function of interdisciplinary percentile across time periods, showing significant increasing trends with respect to interdisciplinarity. Shaded areas represent 95% confidence intervals.

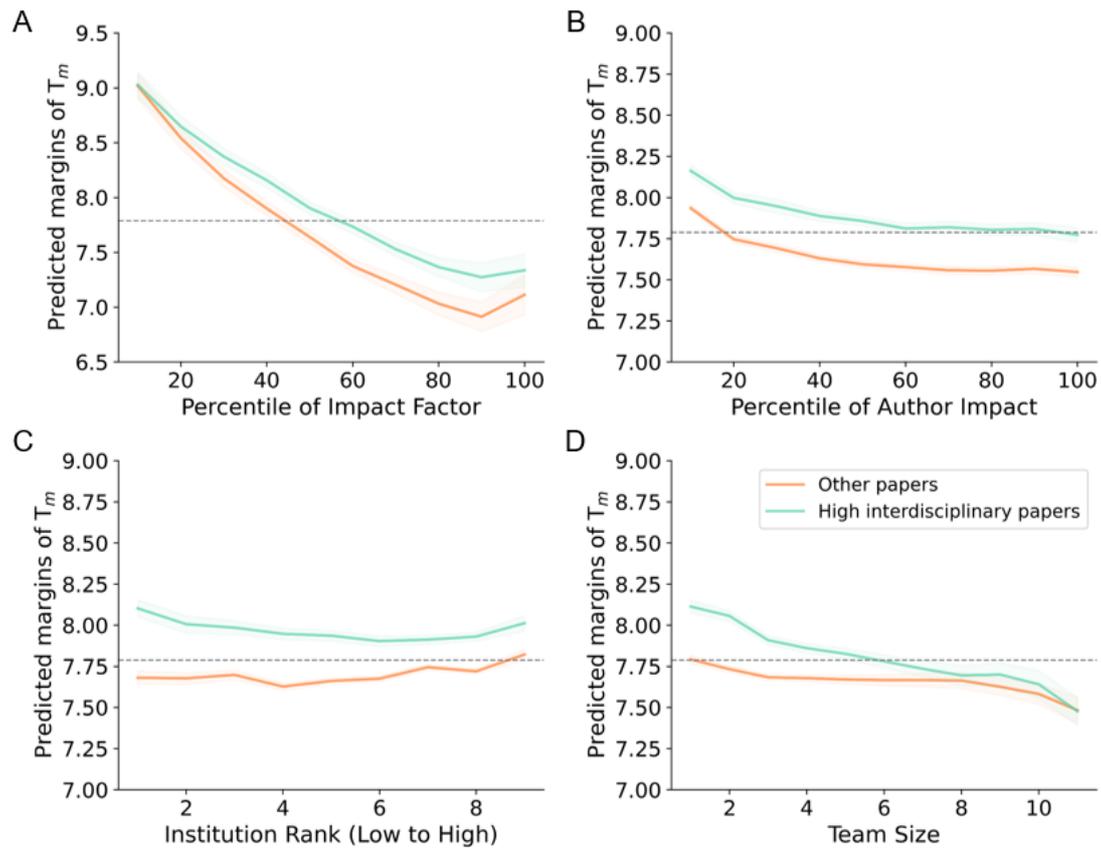

**Figure 4. The Matthew effect and teams.** The predicted margins of $T_m$ conditional on paper interdisciplinarity and (A) percentile of journal impact factor, (B) percentile of author impact, (C) institution rank, and (D) team size, indicating that the Matthew effect and collaboration only play a role in shortening the time to reach the citation peak of individual papers. However, the $T_m$ gap between high interdisciplinary and the other research exists. Shaded areas represent 95% confidence intervals.